\documentclass[]{interact}

\usepackage{epstopdf}
\usepackage{inputenc}
\usepackage[caption=false]{subfig}
\usepackage{dsfont}
\usepackage{amsmath}
\usepackage{xcolor}
\usepackage{siunitx}
\usepackage{stackengine}
\usepackage{graphicx}
\usepackage{setspace}
\usepackage[numbers,sort&compress,merge]{natbib}
\bibpunct[, ]{[}{]}{,}{n}{,}{,}
\renewcommand\bibfont{\fontsize{10}{12}\selectfont}

\theoremstyle{plain}
\theoremstyle{remark}

\DeclareMathOperator*{\argmax}{arg\,max}

\begin{document}

\title{Automated Determination of Hybrid Particle-Field Parameters by Machine Learning}
\author{
    \name{
        Morten Ledum\textsuperscript{a}, Sigbj{\o}rn L{\o}land Bore\textsuperscript{a}, and Michele Cascella\textsuperscript{*,a}\thanks{CONTACT M. Cascella. Email: michele.cascella@kjemi.uio.no}
    }
    \affil{
        \textsuperscript{a}Department of Chemistry and Hylleraas Centre for Quantum Molecular Sciences, University of Oslo, P.O. 1033 Blindern, 0315 Oslo, Norway
    }
}

\maketitle
\begin{abstract}
    The hybrid particle-field molecular dynamics method is an efficient alternative to standard particle-based coarse grained approaches. In this work, we propose an automated protocol for optimisation of the effective parameters that define interaction energy density functional, based on Bayesian optimization. The machine-learning protocol makes use of an arbitrary fitness function defined upon a set of observables of relevance, which are optimally matched by an iterative process. Employing phospholipid bilayers as test systems, we demonstrate that the parameters obtained through our protocol are able to reproduce reference data better than currently employed sets derived by Flory-Huggins models. The optimisation procedure is robust and yields physically sound values.  Moreover, we show that the parameters are satisfactorily transferable among chemically analogous species. Our protocol is general, and does not require heuristic a posteriori rebalancing. Therefore it is particularly suited for optimisation of reliable hybrid particle-field potentials of complex chemical mixtures, and extends the applicability corresponding simulations to all those systems for which calibration of the density functionals may not be done via simple theoretical models. 
\end{abstract}

 
\section{Introduction}
Hybrid particle-field (hPF) simulations are a class of efficient methods that are well adapted for studying very large soft matter systems with molecular resolution.~\cite{Daoulas2006JCP,Muller2011JSP,Milano2009JCP,Vogiatzis2017MACRO} The essence of the hPF methodology is contained in the two terms of the hPF Hamiltonian:
\begin{equation}
     H(\{\mathbf r\})=\sum_m H_0(\{\mathbf r\}_m)+W[\{\phi(\mathbf r)\}].\label{eq:hamiltoniaon}
\end{equation}
Here $H_0$, the Hamiltonian of single molecule $m$, contains the \textit{kinetic energy} and the \textit{intramolecular potential} as defined in standard particle-based potentials, and $W$, the interaction energy functional~\cite{Schmid1998JOP,Kawakatsu2013,Fredrickson2006}  dependent on the density-fields $\phi(\mathbf r)$ of the different particle species, models all \textit{intermolecular interactions}.

Intramolecular forces, by their very nature, only act on a single molecule, while the density-field interactions manifest as a quasi-instantaneous external potential, coupling the motion of the different molecules. The possibility of computing the external potential using particle-mesh routines allows for a very efficient and highly parallel implementation requiring very little communication among processors, resulting in algorithms formally exhibiting strong-scaling~\cite{Zhao2012JCP,Schneider2019CPC}. Very recently, a GPU-based
implementation of the Monte Carlo-based hPF
(\emph{single chain in mean field}) set a new milestone with simulations
of polymer melts composed by 10 billion particles~\cite{Schneider2019CPC}. 

The coupling of hPF to molecular dynamics in efficient parallelised software~\cite{Zhao2012JCP,zhu2013galamost}  has allowed for the application of hPF simulations on both conventional soft polymer
 mixtures and biological systems~\cite{Milano2013PHYSBIO,Soares2017JPCL,Cascella2015CHEMMOD,Marrink2019CHEMREV,bore2018hybrid}. Prominent examples range from nanostructured multiphase materials~\cite{Nicola2016EPJ,Zhao2016NANOSCALE,Munao2018NANOSCALE,Munao2019MACRO} to
organised and disorganised lipid/water mixtures~\cite{Nicolia2012TCA,Nicolia2011JCTC}. Recently, hPF was extended to simulations of polypeptides~\cite{bore2018hybrid}, and to include explicit treatment of electrostatic interactions~\cite{Zhu2016PCCP,kolli2018JCTC,Bore2019JCTC}, the latter opening to the formulation of density functional-based computational predictive models of the complex phase behavior of lipopolysaccharides~\cite{Denicola2020BBA}.

Despite the growing level of maturity reached by hPF simulations, so far relatively little attention has been put into developing systematic protocols for the parameterisation of the interaction energy functional $W$. In particular, the quality of hPF models depends on both the physical model chosen for $W[\phi]$, and on the appropriate calibration of all the numerical parameters it may depend upon. The most commonly employed model for $W$ typically takes the form of:
\begin{equation}
    W\left[\phi(\mathbf{r})\right] = \frac{1}{2\phi_0}\int\mathrm{d}\mathbf{r}\, \left(\sum_{ij} \tilde\chi_{ij}\phi_i(\mathbf{r})\phi_j(\mathbf{r}) + \frac{1}{\kappa}\left(\sum_j \phi_j(\mathbf{r})-\phi_0\right)^2\right), \label{eq:W}
\end{equation}
where the average number density of the system is denoted $\phi_0$, $\kappa$ is a compressibility term which controls the level of fluctuations of the overall density, and the $\tilde \chi_{ij}$ matrix is an energetic parameter that models local mixing energy between species $i,j$ present in the system.

Parameters for the local mixing energy may be derived by different experimental approaches. For example, for simple polymers in a solvent, the $\tilde\chi$-parameter can be obtained from thermometric data~\cite{brandrup1989polymer}. This is however not as easily available when considering hetero-polymeric systems. Another approach is to estimate $\tilde\chi$ by its relationship with the \textit{Hildebrand solubility parameter}~\cite{lindvig2002flory}. However this can be problematic as solubility parameters are often inaccurate~\cite{venkatram2019critical}. Most importantly, for the molecular resolution of hPF models, which often adopt coarse grained (CG) representations in the range of four\textendash ten atoms per bead, factorisation of global experimental data into the individual molecular components may not be trivial. 

A more effective determination of $\tilde\chi$ parameters may be obtained using simple Flory-Huggins (F-H) lattice models:
\begin{equation}\label{eq:fh-bind}
    \tilde \chi_{ij}=-z\left(\epsilon_{ij}-\tfrac 12 \left(\epsilon_{ii}+\epsilon_{jj}\right)\right),
\end{equation}
where $\epsilon_{ij}$ is the mixing energy between species $i$ and $j$, and $z$ is the coordination number, which takes the value of 6 for three-dimensional Cartesian lattices. The mixing energy between two species can be approximated by the two-body interaction energy defined in the potential of the underlying molecular model employed. While this approach has been quite successful so far, there are a few limitations that hamper its general use. Prominently, the F-H model considers contact energies only, sometimes even disregarding entropy contributions to the binding, not  taking into account many-body effects, or long-range interactions. The latter are particularly important, for example, in very polar or charged moieties. 
In practice, F-H parameters provide very good qualitative guesses for the values of $\tilde\chi$. Nonetheless, satisfactory quantitative agreement with reference data, especially in chemically complex systems, usually requires an {\it a posteriori} heuristic fine tuning of at least some of the values of the $\tilde \chi$ matrix~\cite{Nicolia2011JCTC}.

Importantly, even though the first term of the interaction energy in \eqref{eq:W} accounts in principle for the total energy of mixing, in recent times the addition of other terms to the $W$ functional, for example explicitly describing   electrostatics~\cite{Zhu2016PCCP,kolli2018JCTC,Bore2019JCTC} or surface interactions~\cite{sgouros2018mesoscopic,bore2020hybrid}, poses the problem of appropriately factorising such contributions out the mixing $\tilde \chi$ term to avoid  non-physical double-counting. In these cases,  $\tilde\chi$ loses a direct physical meaning, and for this reason it is problematic to define plausible values for $\tilde\chi$ directly from theoretical models.

The hPF interaction energy is globally dependent on a large set of parameters comprising both the $\tilde\chi$ matrix, and any other parameter present in other energy terms eventually employed. Therefore, the determination of a accurate functional $W$ should be addressed as a global optimisation problem. Systematic approaches to parameterisation of ordinary particle-particle potentials in CG force fields, such as force matching~\cite{Izvekov2005JPCB}, Iterative Boltzmann Inversion~\cite{Reith2003JCC} and Inverse Monte Carlo~\cite{Lyubartsev2017CRC}, effectively consider parameterisation as optimisation problems where parameters are chosen to satisfy a given fitness function. For example, Iterative Boltzmann inversion and Inverse Monte Carlo consider a high resolution reference potential of mean force and optimise interaction potentials to reproduce this reference using the CG degrees of freedom. A key observation in such attempts is that the potential of mean force and interactions potentials, due to loss of entropy in the process of CG, most often are significantly different 
Similarly, the $\tilde\chi$ parameter of continuum density-field for polymers has not the direct meaning of a potential of mean force, but rather that of a phenomenological energetic term~\cite{Fredrickson2006}. 

The determination of hPF $\tilde\chi$ force fields parameters poses a particularly challenging optimisation problem. First, these interactions cannot be framed as in a reaction coordinate form; therefore, $\tilde\chi$ parameters cannot be optimised through standard state of the art methods, such as Iterative Boltzmann inversion or Inverse Monte Carlo. Second, the gathering of hPF data does not yield derivatives of the model fitness with respect to the parameters, thereby restricting us to gradient free optimisation. Finally, the $\tilde\chi$-matrix may involve a large parameter space for complex chemical mixtures, thus a general optimisation method needs to be capable of dealing with large dimensional parameter spaces.

Given such constraints, the large family of surrogate (or response surface methodology [RSM]) model based approaches, in which a response surface meta-model is introduced and updated through sequential noisy sampling, provides several possible optimisation techniques. Methods in the literature, of particular relevance, are classical sequential RSM~\cite{amaran2016simulation,myers2016response}, Lipschitz optimisation~\cite{piyavskii1972algorithm,shubert1972sequential}, Trust region methods~\cite{amaran2016simulation,chang2013stochastic}, and Bayesian optimisation~\cite{mockus1978application,kushner1962versatile,pelikan1999boa} (BO). In addition, various random search methods, such as genetic algorithms~\cite{whitley1994genetic,reeves1997genetic}, simulated annealing~\cite{kirkpatrick1983optimization,bertsimas1993simulated,alkhamis1999simulated}, Latin hypercube sampling~\cite{mckay1979comparison}, or straight uniform random sampling, are applicable.

Among the cited methodologies, BO is a versatile scheme for the global optimisation of expensive non-linear black-box functions for which derivatives with respect to the input parameters are hard or impossible to compute~\cite{mockus1978application,kushner1962versatile,pelikan1999boa}. The BO algorithm, developed in the 70s, has in the last decade emerged as a strong solution to derivative-free optimisation of computationally expensive and noisy black-box functions, with powerful performance in many practical applications, especially
within the field of machine learning hyper-parameter optimisation~\cite{snoek2012practical,jones2001taxonomy,azimi2012hybrid,bergstra2011algorithms,swersky2013multi}. 

In this work we present a protocol for the optimization of hPF parameters based on BO. The choice of this methodology is based on its strong theoretical convergence properties when paired with an upper-confidence bound acquisition function~\cite{srinivas2009gaussian}, its simple implementation, and its highly data efficient sampling~\cite{shahriari2015taking}. The effectiveness and robustness of our optimization protocol it tested against uniform random sampling, the simplest possible optimisation strategy, and  previous literature data based on F-H models. 

\section{Materials and methods}  
\subsection{The hybrid particle-field method}
The phase space of a molecular system with total energy \eqref{eq:hamiltoniaon} may be sampled either by Monte Carlo~\cite{Daoulas2006JCP}, or by molecular dynamics (hPF-MD)~\cite{Milano2009JCP}. In this work we employed hPF-MD.  

In hPF-MD, the equations of motion for the independent particles are determined by the presence of an external potential obtained as the functional derivative of $W$. Specifically, the potential acting on each particle species $i$ located at position $\mathbf r$ takes the form~\cite{Milano2009JCP}:
\begin{equation}
    V_i^\text{ext}(\mathbf{r}) = \frac{\delta W\left[\phi(\mathbf{r})\right]}{\delta \phi_i(\mathbf{r})} = \frac{1}{\phi_0}\left(\sum_j \tilde\chi_{ij}\phi_j(\mathbf{r}) + \frac{1}{\kappa}\left(\sum_j \phi_j(\mathbf{r})-\phi_0\right)\right). \label{eq:V}
\end{equation}
In the OCCAM hPF-MD software~\cite{Zhao2012JCP}, which we employ in this paper, the related forces are evaluated via a numerical particle-mesh approach from spatial derivatives of the external potential:
\begin{equation}
    \mathbf F_i^\text{ext}(\mathbf{r}) =-\boldsymbol \nabla V_i(\mathbf{r}) = -\frac{1}{\phi_0}\sum_j \left(\tilde\chi_{ij}+\frac 1\kappa\right)\boldsymbol\nabla \phi_j(\mathbf{r}). \label{eq:F}
\end{equation}
For more details on the computation of the forces, see  ref.~\cite{Milano2009JCP}.

\subsection{hPF force field parameterisation protocol}
\begin{figure}
\includegraphics[width=1\textwidth]{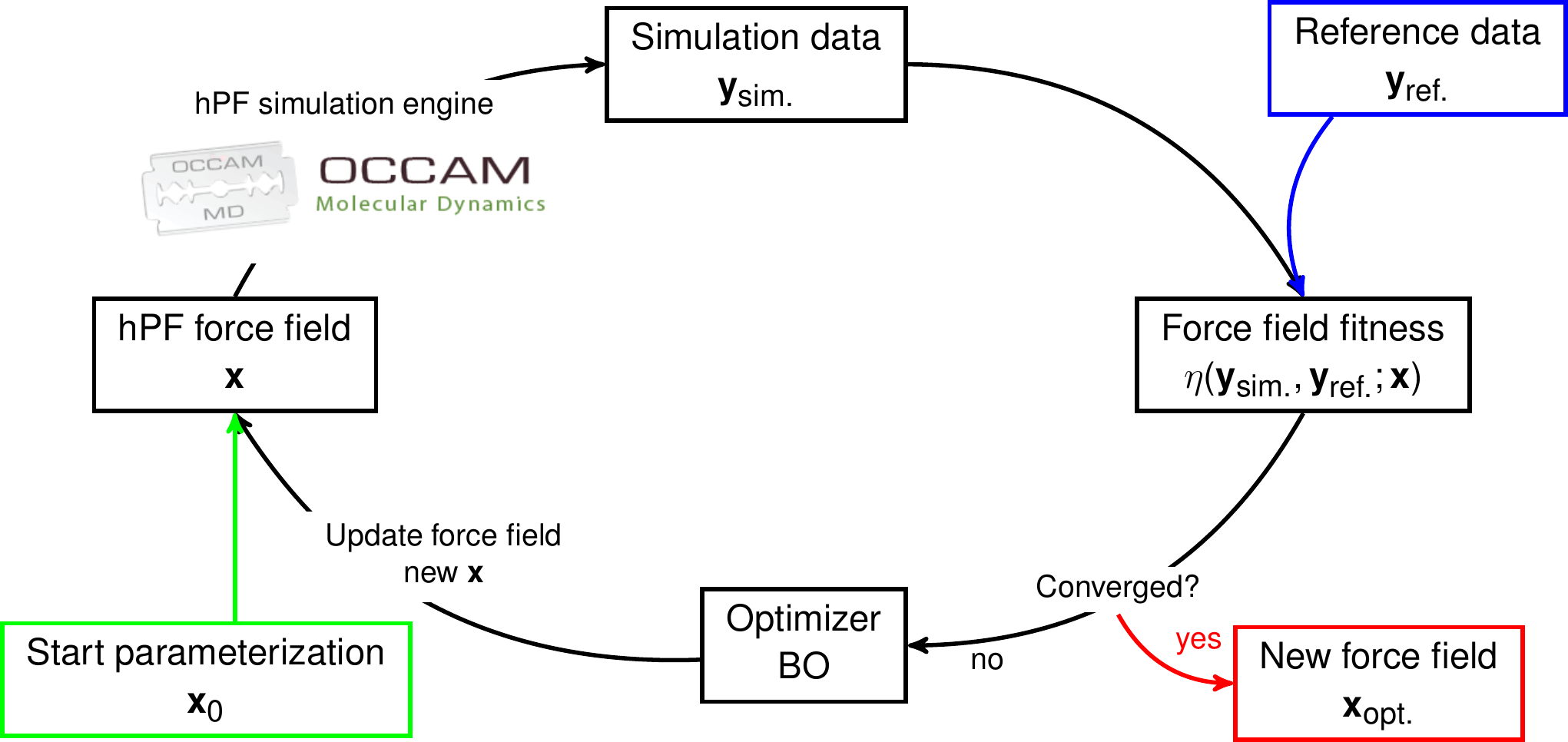}
\caption{Protocol for optimising hPF force fields.}\label{fig:opt}
\end{figure}
To determine hPF force field parameters, we employ a general iterative automated optimisation framework as depicted in Figure~\ref{fig:opt}. Starting from a force field parameter set $\mathbf x$, a hPF trajectory is gathered and analysed giving output data of relevance $\mathbf y_{\text{sim.}}$. The output data is then compared to reference  data $\mathbf y_{\text {ref.}}$, which can be provided by any accurate source, including high(er) resolution simulations or experiment. An objective (or fitness) function $\eta=\eta(\mathbf y_{\text {sim.}}, \mathbf y_{\text {ref.}};\mathbf x)$ assesses the quality of the parameterization, and from the fitness value, the optimiser proposes a new hPF parameter set $\mathbf x$. The full cycle is  automated and is repeated until satisfactory convergence of the fitness is reached, yielding the optimal hPF force field.

In principle, any optimiser that is not dependent on gradient values of the fitness values, can be employed. However, given the potentially large dimension of the parameter space $\mathbf x$ and the computationally expensive simulations needed to gather $\mathbf y_{\text{sim.}}$, it is essential that the optimiser should converge with the fewest possible amount of iterations. Next, simulation data has an element of stochasticity, therefore the optimiser needs to be robust against noise in the fitness values. Finally, we note that computational expensiveness of gathering $\mathbf y_{\text{sim.}}$, makes almost any computational cost of the optimiser itself negligible. 

The protocol we propose makes use of BO, a surrogate based model for solving constrained optimisation problems:
\begin{align}
  \mathbf{x}_\text{opt} = \argmax_{\mathbf{x}\in\mathcal{X}} \eta(\mathbf{x}).
\end{align}
The space of possible parameter configurations $\mathcal{X}$ is usually a compact subset of $\mathds{R}$ and the objective function $\eta$ is in general unknown, non-convex, multimodal, and only accessible through (computationally expensive) pointwise noisy sampling. In the BO algorithm, a Gaussian process (GP) function prior is placed on the underlying true objective and updated via Bayesian posterior updating (Bayes' rule) by sequential probing of $\eta$~\cite{shahriari2015taking}. In this way, a probabilistic response surface is built which represents, at each iteration, the model's
beliefs about the objective ($\mu$) and how confident the model is at each point in $\mathcal{X}$ ($\sigma$). BO achieves high efficiency in the sampling of the parameter space by leveraging both $\mu$ and $\sigma$ in an \textit{acquisition function} (AF), $a(\mathbf{x})=a(\mu(\mathbf{x}),\sigma(\mathbf{x}))$. Often, the AF contains a parameter $\beta$ which governs the trade-off between
exploration (sampling areas in $\mathcal{X}$ where the uncertainty is high) and exploitation (sampling areas where \textit{good} $\mathbf{x}$ are known to be located). The AF guides the sampling by picking points $\mathbf{x}\in\mathcal{X}$ to explore according to a strategy for improving upon the currently best found $\mathbf{x}$.

The GP prior is a multivariate Gaussian distribution over functions,
uniquely defined by a covariance kernel $\Sigma_0$ and a mean function $\mu_0$. The kernel function induces a metric on $\mathcal{X}$ which defines a measure of the distance (similarity) between points $\mathbf{x}$ and $\mathbf{x}'$. The choice of a specific such $\Sigma_0$ represents a priori assumptions about the structure of the underlying true objective.

Often, one or more hyper-parameters in the covariance kernel have to be specified. It is customary to fix the values of these parameters by the \textit{marginal likelihood} of the model, given the observed data.
Marginalizing out the true noise-free objective function gives the likelihood of the model hyper-parameters. For GPs, the log marginal likelihood integral in question is analytically tractable, and may be easily maximised to determine the optimal kernel hyper-parameters.

\subsection{Test case: Phospholipid model for bilayers}
\begin{figure}
    \centering
    \includegraphics{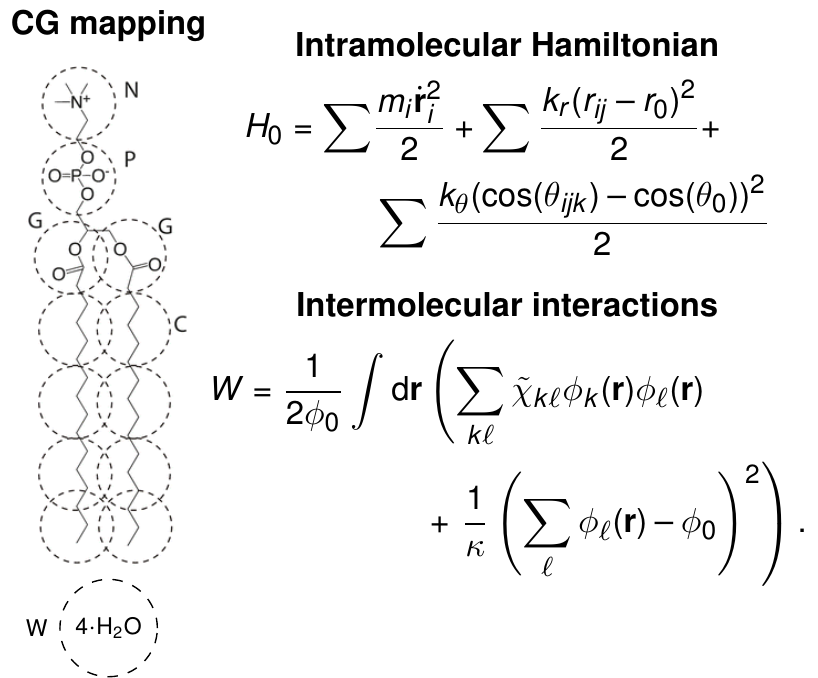}
  \caption{Summary of the hPF phospholipid model. {\it Left:} CG representation of the DPPC phospholipid and solvent. Right: Outline of the two terms in the hPF Hamiltonian.}\label{fig:cg-rep}
\end{figure}

As test case we consider a hPF-MD model for fully-saturated phospholipid bilayers, using in particular four variants characterised by different lengths of the fatty tail dipalmitoylphosphatidylcholine (DPPC), dimyristoylphosphatidylcholine (DMPC), distearoylphosphatidylcholine (DSPC), and mono-unsaturated dioleoylphosphatidylcholine (DOPC). For direct comparison, we use the same mapping of the model developed by De Nicola et al.~\cite{Nicolia2011JCTC} (Figure~\ref{fig:cg-rep}), which employs a MARTINI CG representation of the phospholipids \cite{wassenaar2015computational} and explicit solvent. 

In this work, we limit our analysis to the optimisation of the  $\tilde\chi_{ij}$ matrix, while the bonded terms and the compressibility $\kappa$ are kept the same in the model of De Nicola~\cite{Nicolia2011JCTC}.

A ($13\times13\times\SI{14}{nm^3}$) simulation box containing 528 DPPC lipids and 14000 water beads  is employed. Each simulation in the optimisation look lasts \SI{20}{ns}. The hPF simulations were run using OCCAM.\footnote{http://www.occammd.org/} The simulations are performed under the $NVT$ ensemble, using the Andersen thermostat with a coupling time \SI{0.1}{ps} and collision frequency \SI{7.0}{ps^{-1}}. A time step of \SI{0.03}{ps} was used. The particle-mesh routines for particle-field forces in OCCAM employed a grid size of \SI{0.58}{nm} ($1.25$ times the bond length used) and an update period of \SI{0.3}{ps} (10 time steps). hPF-MD simulations are performed at a temperatures of \SI{335}{\kelvin}, \SI{325}{\kelvin}, and \SI{303}{\kelvin} for DSPC, DPPC, and DMPC/DOPC, respectively.  

To evaluate the fitness of the model we consider  electron density profiles ($\varphi$) of the different species, compared to those obtained from reference CG simulations using the MARTINI force field. 
This choice is made to have the best assessment of the quality of the BO procedure as compared to F-H. For optimal determination of hPF parameters for phospholipids, more accurate all-atom models may be eventually employed.  

The fitness $\eta(\varphi;\tilde{\chi})$ is defined as the average mean squared error over the electron densities of the different species $k$:
\begin{equation}
    \eta(\varphi;\tilde{\chi}) = \frac{1}{nn_k}\sum_{k=1}^{n_k}\sum_{i=1}^n \big\vert\varphi_i^k - \hat{\varphi}_i^k\big\vert^2, \label{eq:mse}
\end{equation}
with $\varphi_i^k$ being the electron density of species $k$ at a position $z_i=2i\ell/n - \ell$ along the bilayer normal. The density profiles are computed relative to the center of mass of all carbon type beads in the simulation, which is taken to be the center of the bilayer. $\hat{\varphi}_i^k$ indicates the reference density to be matched (in our case the MARTINI simulation results). The total number of different particle species is denoted $n_k$, while $n$ is the number of bins in the chosen density histogram. For a better of comparison with F-H data~\cite{Nicolia2011JCTC}, the absolute deviations $S_k$ are also reported: 
\begin{align}
    S_k(\varphi;\tilde{\chi}) = \frac{1}{n}\sum_{i=1}^n \vert \varphi_i^k - \hat{\varphi}^k_i\vert.
\end{align}
In addition, $S_\text{p}$, the mean percentage error relative to the average electron density  $\varphi_0$ over the full histogram across all species, is reported: 
\begin{align}
    S_\text{p}(\varphi;\tilde{\chi}) = \frac{1}{\varphi_0nn_k}\sum_{k=1}^{n_k}\sum_{i=1}^n \vert \varphi_i^k - \hat{\varphi}_i^k \vert,
\end{align}

To avoid potential {\it cold-start} problems, each optimisation run is started with $2d$ ($d$ being the dimension of the parameter space) randomly sampled points. After the initial random sampling period, new points to be probed are selected according to the maximum of the UCB acquisition function~\cite{srinivas2009gaussian} (Figure~\ref{fig:opt}). All $\tilde{\chi}$ parameters are constrained to the values used by De Nicola $\pm\SI{10}{\kilo\joule\per\mol}$~\cite{Nicolia2011JCTC}. The exploration/exploitation trade-off parameter in the acquisition function is set to $\beta=2$, favoring exploration of the large parameter space. The Gaussian process underlying the BO uses a Mat{\' e}rn covariance kernel~\cite{matern2013spatial,stein2012interpolation} with smoothing parameter $\nu=5/2$, and a constant zero mean function. In addition, a diagonal white noise kernel is added to account for the noisy sampling.

\section{Results and discussion}

\subsection{Optimisation of hPF parameters for DPPC}
DPPC was used as prototypic test systems to assess the effectiveness of BO for the determination of hPF $\tilde\chi$ parameters. The choice of such system was determined both by the presence of a relatively complex chemical structure, and by the existence of a vast reference literature, including  experimental~\cite{Petrache2000BJ,Waheed2009BJ,Nagle2000BBA} and computer simulations~\cite{Lindahl2000JCP,Levine2014JACS}, as well as hPF models~\cite{Nicolia2011JCTC,Nicolia2012TCA}.

\captionsetup[subfigure]{labelformat=empty,position=top,textfont=small}
        \begin{figure}
            \centering
            \subfloat[Flory-Huggins]{%
                \hspace{-0.45cm}\resizebox*{4.9cm}{!}{\includegraphics{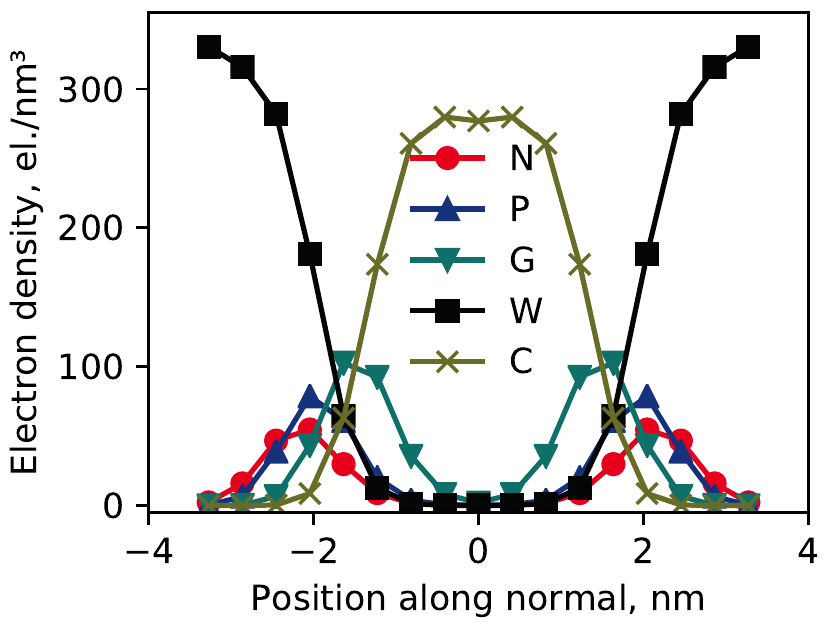}}%
            }
            \subfloat[\hspace{0.45cm}MARTINI]{%
                \hspace{0.45cm}\resizebox*{4.9cm}{!}{\includegraphics{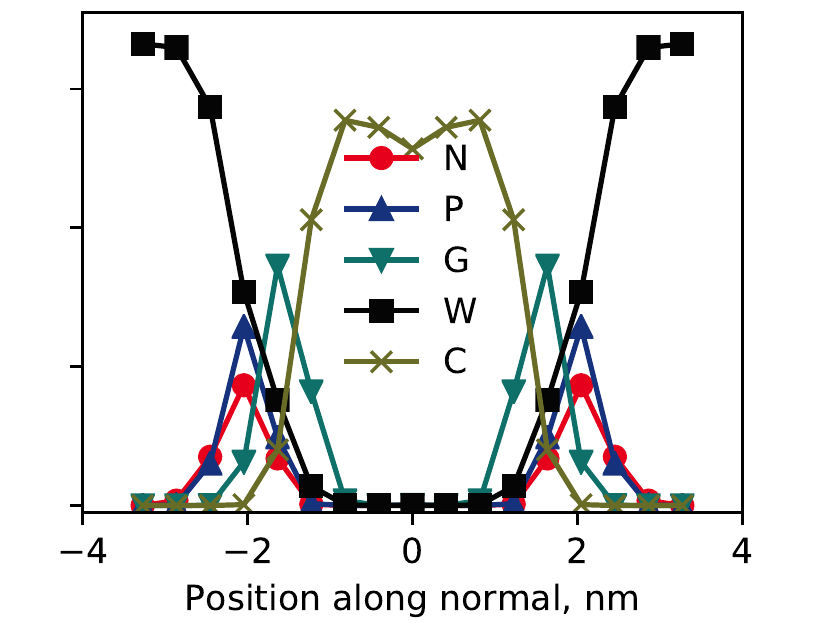}}%
            }
            \subfloat[Bayesian optimisation]{%
                \resizebox*{4.9cm}{!}{\includegraphics{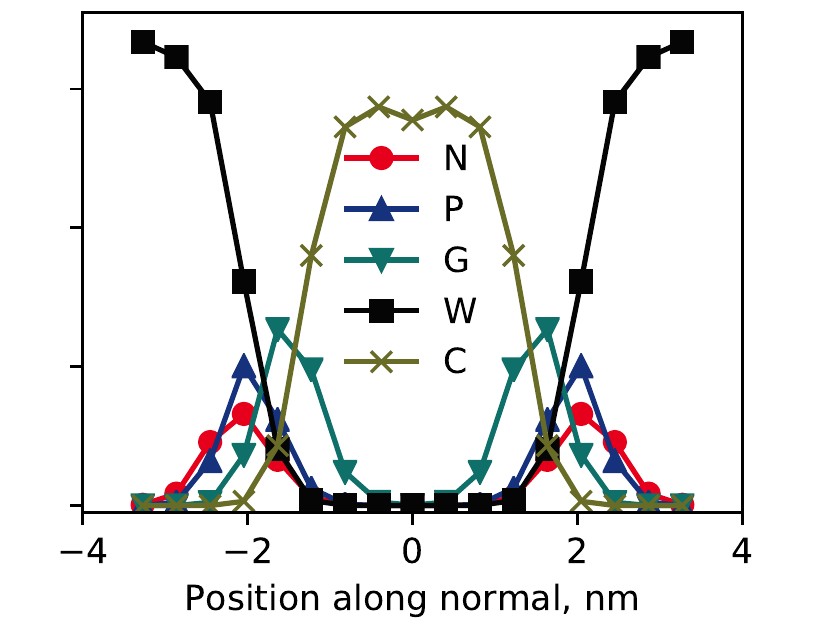}}%
            }\\
            \subfloat{%
                \resizebox*{4.9cm}{!}{\includegraphics[clip,trim=5cm 5cm 5cm 5cm]{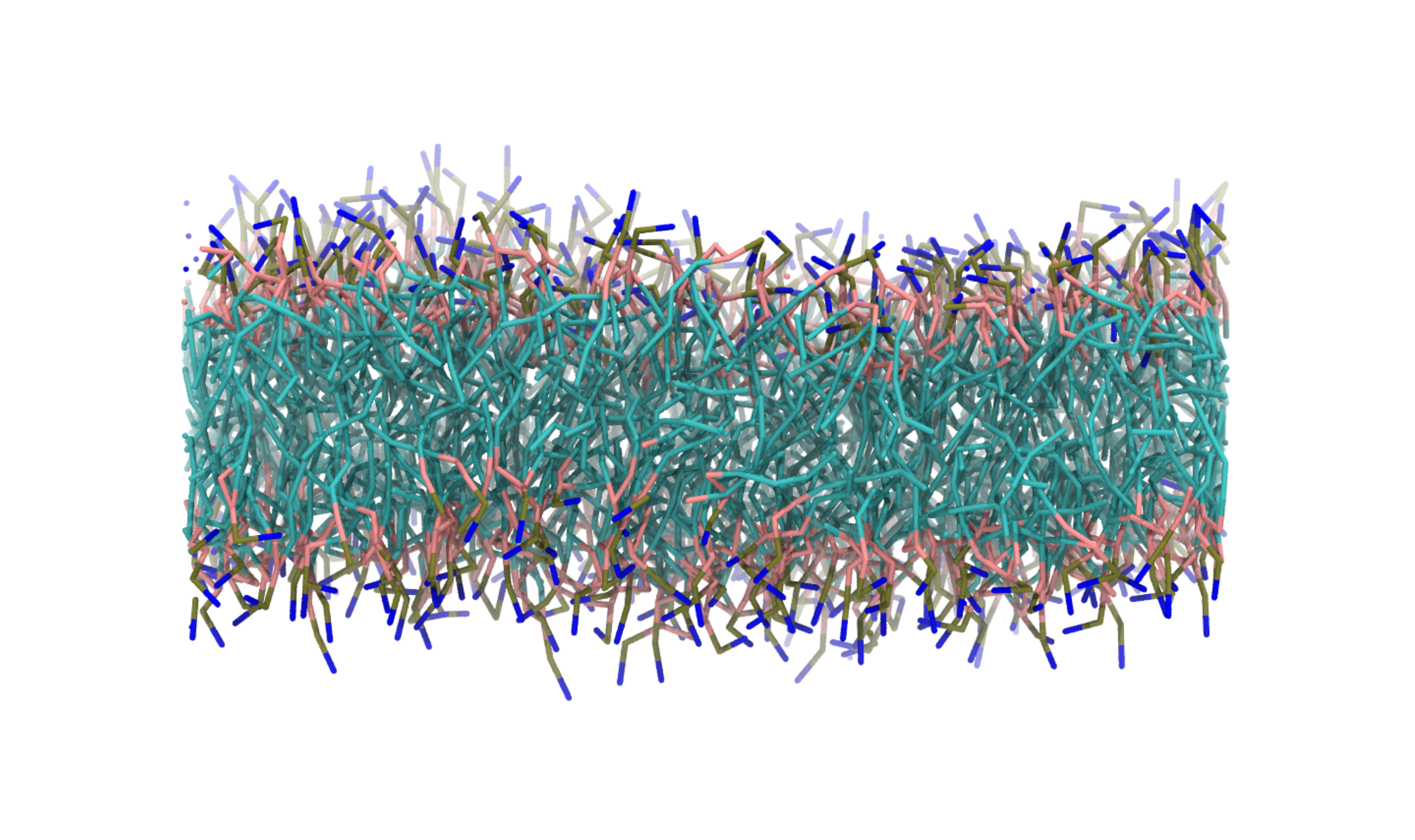}}%
            }
            \subfloat{%
                \resizebox*{4.9cm}{!}{\includegraphics[clip,trim=5cm 5cm 5cm 5cm]{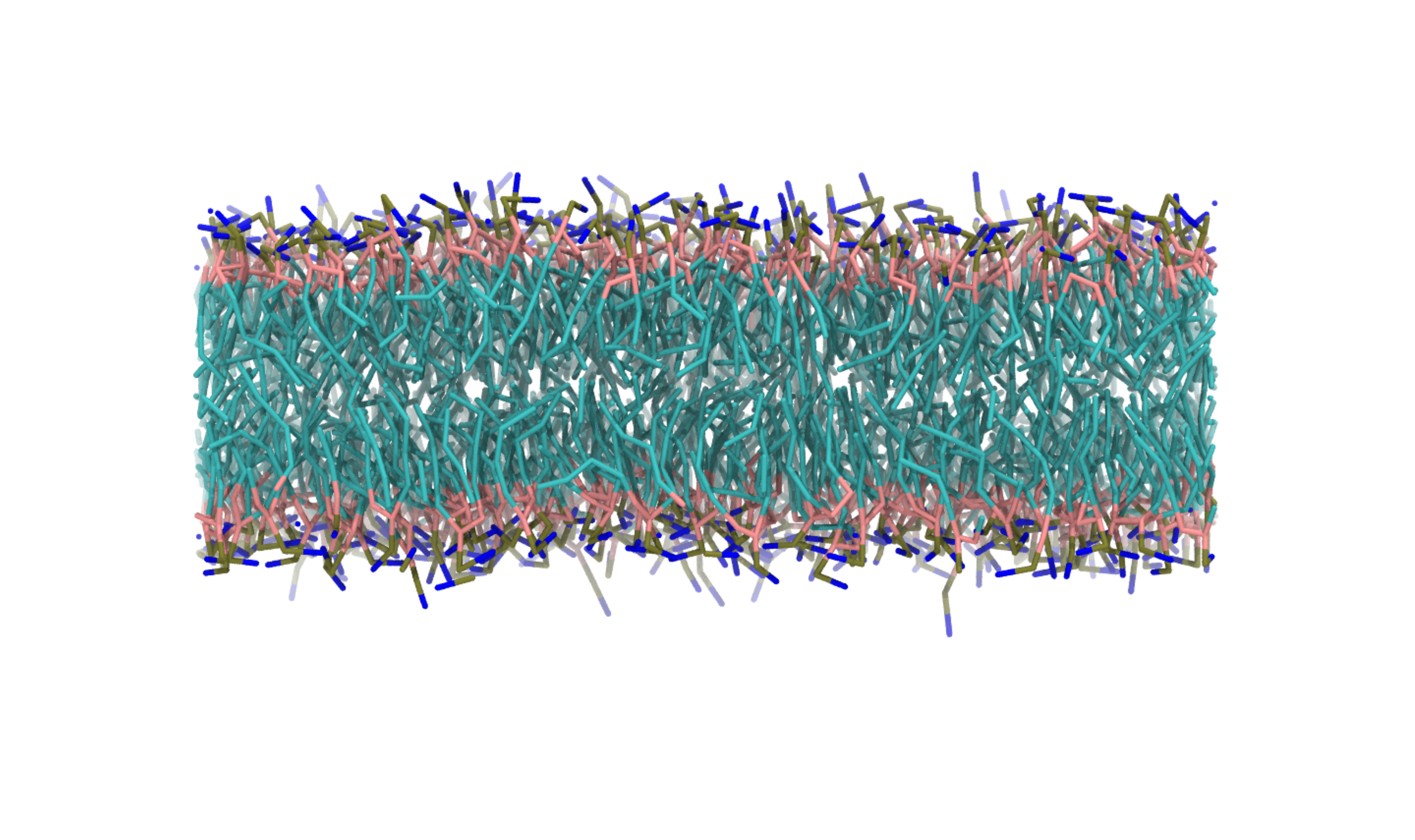}}%
            }
            \subfloat{%
                \resizebox*{4.9cm}{!}{\includegraphics[clip,trim=5cm 5cm 5cm 5cm]{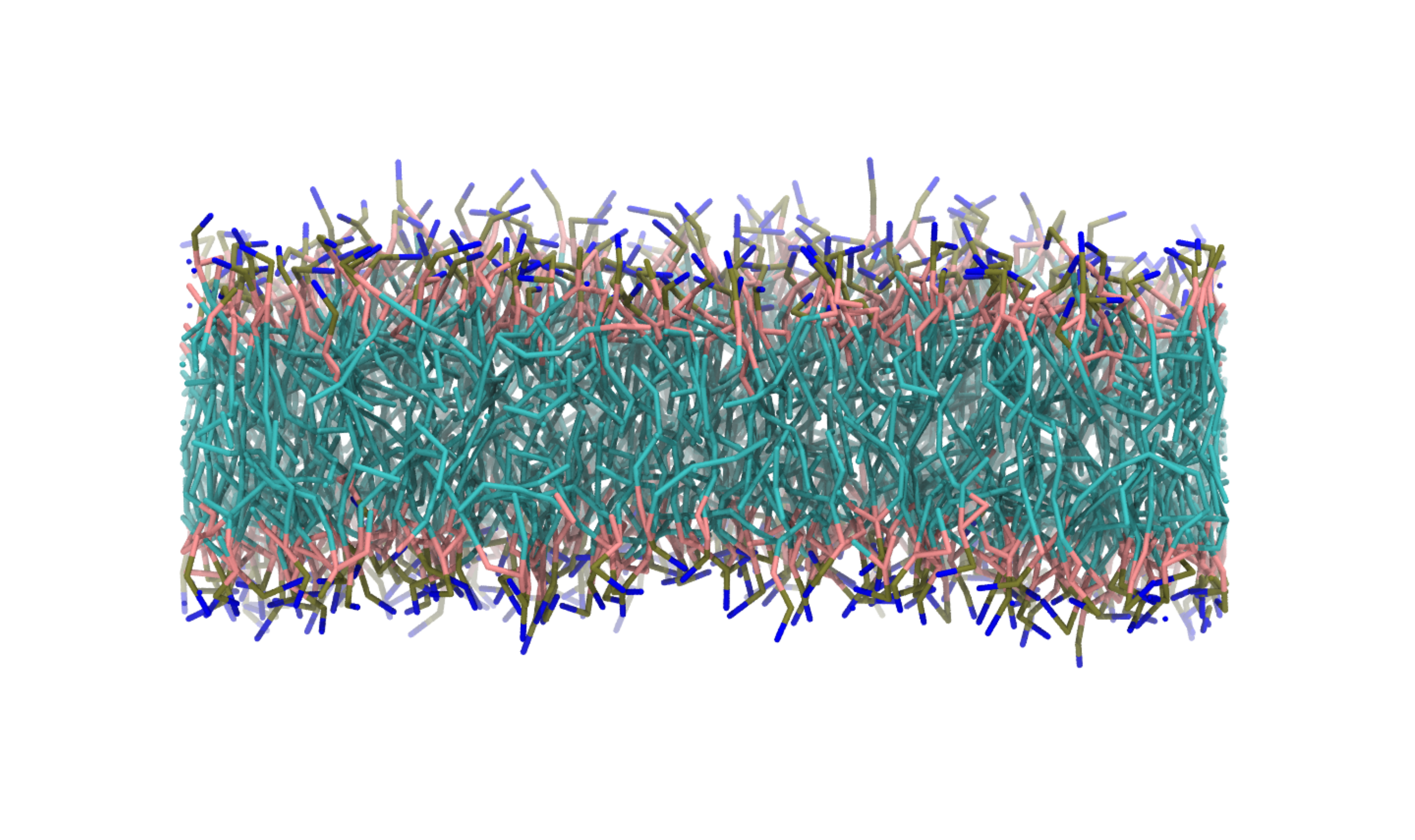}}%
            }\\
            \footnotesize\begin{tabular}{lrrrrrr}
    \toprule
          & N & P & G & C & W & average \\
    \midrule
BO (this work)   & 4.65 & 4.10  & 6.52  & 7.26  & 8.91  & 6.29 (1.71\%) \\
F-H~\cite{Nicolia2011JCTC} & 9.29 & 12.19 & 20.51 & 12.23 & 12.82 & 13.41 (4.10\%) \\
    \bottomrule
            \end{tabular}
            \caption{Density profiles and representative membrane snapshots from hPF-MD simulations of a DPPC bilayer using $\tilde\chi_\text{F-H}$ parameters~\cite{Nicolia2011JCTC} ({\it left}), particle-based simulations using the MARTINI CG force field ({\it centre}), and  hPF-MD $\tilde\chi_\text{BO}$ parameters ({\it right}). The table presents absolute deviations $S_k$ in the density profiles between the F-H and BO parameter simulations, and the reference MARTINI profile. Percentage deviations $S_\text{p}$ are given in parenthesis. $S_k$ values are given in el./\si{\nano\meter\cubed}.}
            \label{fig:dppc-combo}
        \end{figure}

Figure~\ref{fig:dppc-combo} reports the density profiles for hPF simulations of DPPC after BO of the $\tilde\chi$ parameters  with respect to the mean-square-error objective function, computed between the  hPF and reference MARTINI density profiles
\eqref{eq:mse}. The density profiles for all the bead types match well those of the reference,  with $S_\text{k}$ values smaller than \SI{7.3}{el./nm^3} for all lipid beads, and with a $S_\text{p}$ value less than \SI{2}{\percent}.

Previously published hPF-MD models for phospholipids are based on the MARTINI CG mapping~\cite{marrink2004coarse,Marrink2007JCPB,wassenaar2015computational} and employ a $\tilde\chi$ matrix based on the F-H model \eqref{eq:fh-bind}. F-H parameters are extracted from the corresponding Lennard-Jones binding energies of the MARTINI force field ($\tilde{\chi}_\text{F-H}$ hereafter~\cite{Nicolia2011JCTC}). As noted in the original work, using the lateral density profile as the benchmark property, heuristic adjustment of the $\tilde\chi$ parameter between C and W beads was required to improve the stability and overall structure of the bilayer. 
Overall, the F-H parameter set produces a satisfactory organisation of the lipid bilayer (Figure~\ref{fig:dppc-combo}), evidenced by a very good qualitative agreement of the lateral density profiles for the different moieties compared to reference CG simulations using the MARTINI force field. Nonetheless, the hPF/F-H density profiles are characterized by a $S_\text{p}$ of about $4-\SI{5}{\percent}$ for the different lipids~\cite{Nicolia2011JCTC}, and larger values of $S_\text{k}$, reaching a maximum of \SI{20.51}{el./nm^3} for the G bead. Comparison of $S_\text{k}$ and $S_\text{p}$ values indicates that BO provides a substantial improvement compared to F-H. 

Given the use of theoretical models for the derivation of $\tilde\chi$, it has been hard so far to discern the origin of any discrepancies from reference data between intrinsic approximations of the hPF method, or the use of non optimal parameter sets. 
In particular, broader density profiles in lipids  were usually understood as a consequence of the intrinsic softness of the field interactions~\cite{Zhu2016PCCP}. In fact, using BO parameters, there is an appreciable sharpening of the distributions for all the beads, even though the peaks remain broader than CG simulations based on pair-interactions (Figure~\ref{fig:dppc-combo}). This is of particular interest, as it demonstrates that indeed, in phospholipids, the F-H parameterisation is accurate enough to capture the physics of the hPF model; nonetheless, there is still space for significant quantitative improvement by a global optimisation approach.
        
\subsection{Feature importance}
Data in Figure~\ref{fig:dppc-combo}  show how the performance of the F-H parameters is not equal for all the moieties present in the system. In particular, F-H is better at reproducing the distributions of the lipid head and tails, while the density profile of the glycerol groups (G beads) appears too broad. The physical reason for such discrepancy may be attributed to the fact that glycerol floats at the interface between the phase-separated water and lipid fatty tails. Therefore, its distribution depends more than the others on a delicate balance among all the terms in $W$. This effect may be difficult to reproduce adopting an independent parameterisation of the individual elements of the  $\tilde\chi$ mixing energy matrix. On the contrary, the BO approach appears  better suited to take into account all competing interaction, producing more balanced $\tilde\chi$ values.
        
\begin{figure}
    \centering
    \resizebox*{6cm}{!}{\includegraphics{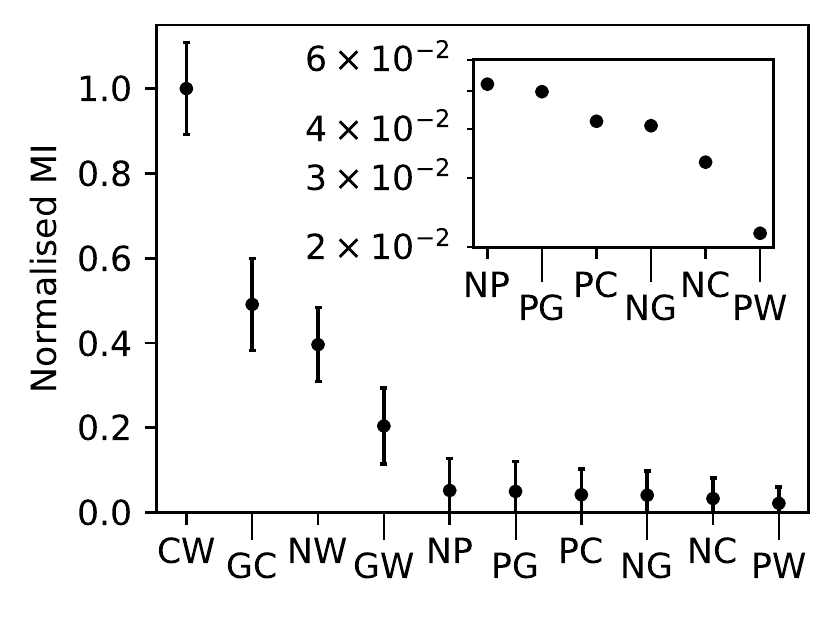}}%
    \hspace{20pt}\resizebox*{6cm}{!}{\includegraphics{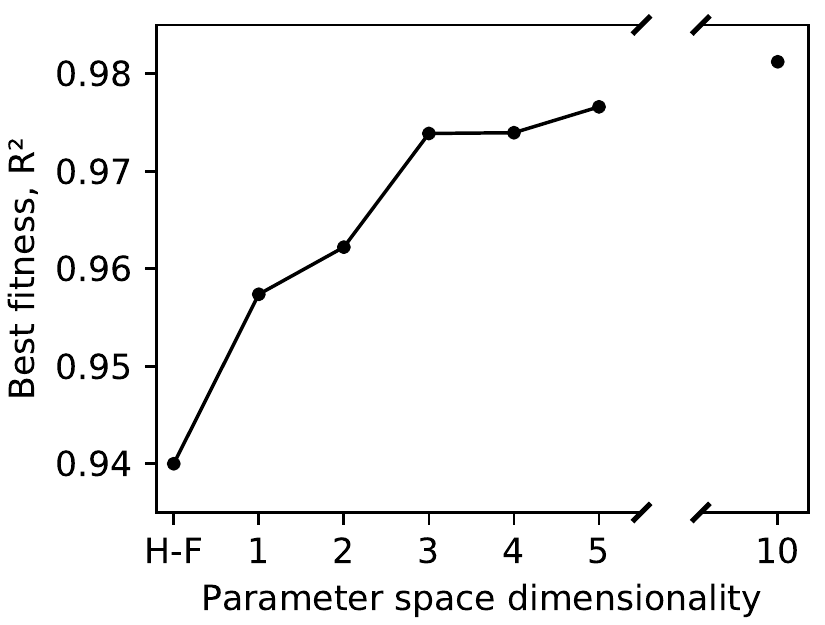}}%
    \caption{{\it Left:} Feature importance as ranked by the mutual information measure between the fitness and the individual $\tilde{\chi}_{ij}$ parameters, for hPF-MD simulations of a DPPC bilayer with randomly sampled $\tilde{\chi}$ matrices. Presented values are normalised relative to the most important parameter ($\tilde{\chi}_\text{CW}$) (arbitrary units). Inset details the low relative MI values found for the last six matrix elements (error bars omitted). {\it Right:} Best fitness achieved (here using the average coefficient of determination, $R^2$, across all bead species) for each dimension of the parameter space subspace used in hPF-MD BO protocol runs on the DPPC bilayer system. $\tilde{\chi}$ parameters are included in order of decreasing feature importance ( left).
    } \label{fig:feature-importance}
\end{figure}

        \captionsetup[subfigure]{labelformat=empty,position=top,textfont=small}
        \begin{figure}
            \centering
            \subfloat[\hspace{22pt}4 parameters]{%
                \resizebox*{6cm}{!}{\includegraphics{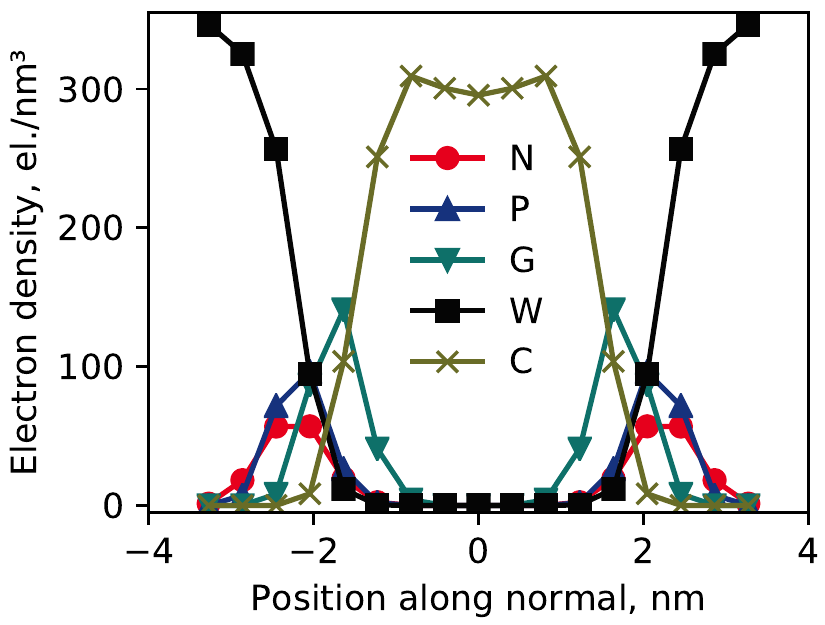}}%
            }%
            \hspace{10pt}\subfloat[10 parameters]{%
                \resizebox*{6cm}{!}{\includegraphics{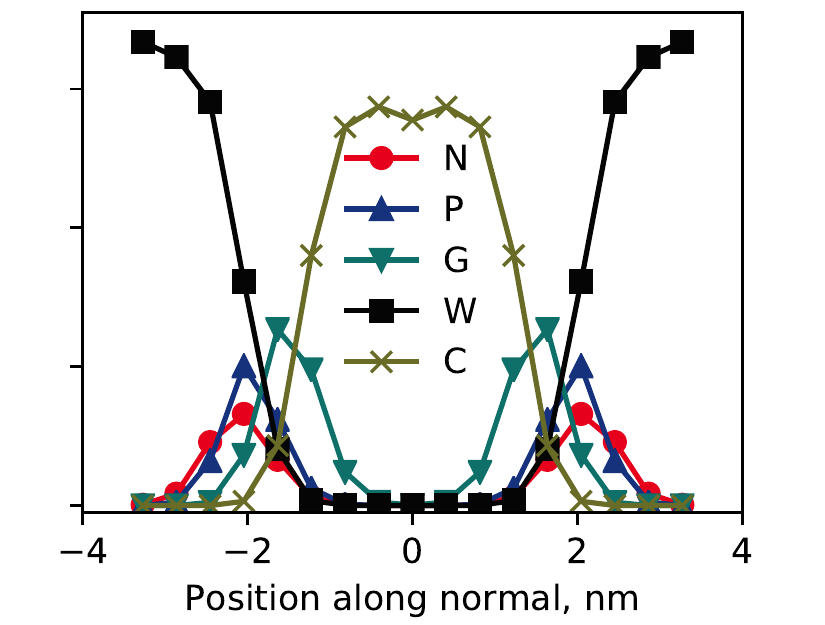}}%
            }\\[10pt]
            {\fontsize{8}{9}\selectfont
            \begin{tabular}{lrrrrr}
                \toprule
                                   & \multicolumn{5}{c}{parameter space dimensionality} \\
                \cmidrule(lr){2-6}
                         & 10 & 4 & 3 & 2 & 1 \\
                \midrule
C\textendash W & 42.24 &  43.68                   & 43.63                  & 42.09                  &  38.16 \\
G\textendash C & 10.47 &  14.00                   & 15.33                  & 14.69                  & \textcolor{red}{ 6.30}\\
N\textendash W & -3.77 &   1.55                   &  1.82                  & \textcolor{red}{-8.10} & \textcolor{red}{-8.10} \\
G\textendash W &  4.53 &   3.02                   & \textcolor{red}{ 4.50} & \textcolor{red}{ 4.50} & \textcolor{red}{ 4.50} \\
N\textendash P & -9.34 &  \textcolor{red}{-1.50}  & \textcolor{red}{-1.50} & \textcolor{red}{-1.50} & \textcolor{red}{-1.50} \\
P\textendash G &  8.04 &  \textcolor{red}{ 4.50}  & \textcolor{red}{ 4.50} & \textcolor{red}{ 4.50} & \textcolor{red}{ 4.50} \\
N\textendash G &  1.97 &  \textcolor{red}{ 6.30}  & \textcolor{red}{ 6.30} & \textcolor{red}{ 6.30} & \textcolor{red}{ 6.30} \\
P\textendash C & 14.72 &  \textcolor{red}{13.50}  & \textcolor{red}{13.50} & \textcolor{red}{13.50} & \textcolor{red}{13.50} \\
P\textendash W & -1.51 &  \textcolor{red}{-3.60}  & \textcolor{red}{-3.60} & \textcolor{red}{-3.60} & \textcolor{red}{-3.60} \\
N\textendash C & 13.56 &  \textcolor{red}{ 9.00}  & \textcolor{red}{ 9.00} & \textcolor{red}{ 9.00} & \textcolor{red}{ 9.00} \\[5pt]
$S_\text{p}$          & 1.71\% & 1.96\% & 2.25\%  & 2.29\%  & 2.32\% \\ 
                \bottomrule
            \end{tabular}}
            \caption{\textit{Top:} Density profiles for hPF-MD DPPC bilayer simulations ran with Bayesian optimised parameter sets with four (left) and ten (right) included $\tilde{\chi}$ parameters. The four-parameter simulation uses $\tilde{\chi}_\text{F-H}$ values for all but the $\tilde{\chi}$ matrix elements with the highest feature importance, namely $\tilde{\chi}_\text{NW}$, $\tilde{\chi}_\text{CW}$, $\tilde{\chi}_\text{GW}$, and $\tilde{\chi}_\text{GC}$, c.f.\ column three of the table (bottom). \textit{Bottom:} Resulting $\tilde{\chi}$ matrices from the BO protocol applied to hPF-MD simulations of a DPPC bilayer. Results reported for selected subspaces of the full 10-dimensional parameter space, with $\tilde{\chi}_{ij}$s shown in red being fixed and not part of the optimisation run. All $\tilde{\chi}$ values given in \si{\kilo\joule\per\mol}. Mean percentage errors, $S_\text{p}$, associated with each set of optimised parameters is given in the last row.}
            \label{fig:4-parameters-profile}
        \end{figure}

The uneven error in the F-H distributions suggests that the hPF model is not equally robust with respect to variations of the different $\tilde\chi$ terms. To verify this hypothesis, we calculated the correlation (mutual information, MI) between input $\tilde\chi$ parameters and resulting fitness. The MI between two continuous random variables $X$ and $Y$ with probability density functions $f_X$ and $f_Y$ (and joint PDF $f_{X,Y}$) is 
\begin{align}
    I(X;Y)=-\int_\mathcal{X}\int_\mathcal{Y}\mathrm{d}x\mathrm{d}y\, f_{X,Y}(x,y)\log \frac{f_{X,Y}(x,y)}{f_X(x)f_Y(x)},
\end{align} 
and can be understood as the reduction in uncertainty about the values of $Y$, once $X$ is revealed~\cite{frenay2013mutual}. The MI between any input parameter $\tilde{\chi}_{kj}$ and the resulting fitness $\eta(\phi;\tilde{\chi})$, thus yields a measure of the feature importance for the full parameter space.

Figure~\ref{fig:feature-importance} shows the relative feature importance of the different $\tilde{\chi}_{ij}$ parameters, as well as optimisation results from BO runs which only include the most important ones. The fitness is here represented by the average coefficient of determination, $R^2$, over the density profiles of all the different beads,
\begin{align}
    \eta^{R^2}(\varphi;\tilde{\chi})=\frac{1}{n_k}\sum_{k=1}^{n_k}R^2(\varphi^k, \hat{\varphi}^k).
\end{align}
Evidently, a subset of just four parameters carry the majority of the feature importance, meaning optimising only these four, keeping the others at their F-H model value, yields results comparable to the ones obtained after an optimisation over the full 10-dimensional parameter space (Figures~\ref{fig:feature-importance} and \ref{fig:4-parameters-profile}). The four relevant parameters have a clear physical meaning, as they are the main determinants for the hydrophilic/hydrophobic character of the polar heads and the fatty tails, respectively ($\tilde{\chi}_\text{NW},\tilde{\chi}_\text{CW}$), and for the amphipathic behaviour of glycerol  ($\tilde{\chi}_\text{GW},\tilde{\chi}_\text{GC}$).

\subsection{Transferability of BO-hPF parameters.}
Table~\ref{tab:parameters-DPPC-opt} reports the parameter sets obtained by BO for DPPC compared to those obtained for two other saturated phospholipids differing in the length of fatty acid chains (DSPC, DMPC), and one unsaturated lipid (DOPC). Overall, the most relevant four $\tilde\chi$ matrix elements do not differ significantly from DPPC to DSPC. The less hydrophobic character of the C bead in DOPC may be attributed to the presence of the unsaturated moiety. We remark that for sake of simplicity, the C=C bond was represented by a different bead type, consistent with the MARTINI mapping, and all $\tilde\chi$ parameters involving that were kept at the reference F-H values~\cite{Nicolia2011JCTC}.

\begin{table}
\tbl{Optimised $\tilde\chi$-matrix parameters found by the BO scheme for hPF-MD simulations of DPPC, DMPC, DSPC, and DOPC bilayer systems. The $\tilde{\chi}$ matrix elements are given in order of decreasing feature importance. Reference parameters are the Flory-Huggins ($\tilde{\chi}_\text{F-H}$) parameters used in~\cite{Nicolia2011JCTC}. All values given in units of \si{\kilo\joule\per\mol}.}
    {\begin{tabular}{lrrrrr}
        \toprule
                          & & \multicolumn{4}{c}{optimised with BO} \\
        \cmidrule(lr){3-6}
            & ref~\cite{Nicolia2011JCTC} & DPPC & DMPC & DSPC & DOPC \\
        \midrule
            C\textendash W & 33.75 & 42.24 &  41.20  & 40.15 & 35.00  \\
            G\textendash C &  6.30 & 10.47 &  13.78  & 14.65 & 14.61  \\
            N\textendash W & -8.10 & -3.77 &  -2.58  & -3.02 & -2.46  \\
            G\textendash W &  4.50 &  4.53 &   5.91  &  4.71 &  9.07  \\
            N\textendash P & -1.50 & -9.34 &  -4.34  & -5.91 & -3.40  \\
            P\textendash G &  4.50 &  8.04 &   5.26  &  7.25 &  8.45  \\
            N\textendash G &  6.30 &  1.97 &   3.37  &  2.99 &  4.92  \\
            P\textendash C & 13.50 & 14.72 &  19.72  & 16.16 &  12.52 \\
            P\textendash W & -3.60 & -1.51 &  -1.26  & -2.17 & -1.27  \\
            N\textendash C &  9.00 & 13.56 &  12.71  & 10.56 & 14.39  \\   
        \bottomrule
    \end{tabular}}
    \label{tab:parameters-DPPC-opt}
\end{table}

        \begin{table}[ht]
\tbl{Mean absolute deviations in electron density, $S_k$, with respect to the MARTINI reference density for the different lipids simulated with Bayesian optimised parameter sets on the different phospholipids (relative percentage deviations $S_\text{p}$ in parenthesis). Comparison with data from De Nicola, using the baseline $\tilde{\chi}_\text{F-H}$ parameter set~\cite{Nicolia2011JCTC}. All $S_k$ values given el./\si{\nano\meter\cubed}.}
    {\begin{tabular}{lrrrrrr}
\toprule
  & \multicolumn{6}{c}{DPPC} \\
  \cmidrule(lr){2-7}
  & N & P & G & C & W & average \\
\midrule
DPPC-optimised                & 4.65 & 4.10  & 6.52  & 7.26  & 8.91  & 6.29 (1.71\%) \\
DMPC-optimised                & 4.45 & 4.05  & 6.44  & 8.37  & 8.36  & 6.34 (1.97\%) \\
DOPC-optimised                & 5.39 & 8.15  & 12.05 & 9.81  & 6.51  & 8.38 (2.61\%) \\
DSPC-optimised                & 5.02 & 6.08  & 8.71  & 9.40  & 9.63  & 7.76 (2.40\%) \\
reference~\cite{Nicolia2011JCTC} & 9.29 & 12.19 & 20.51 & 12.23 & 12.82 & 13.41 (4.10\%) \\[10pt]
  & \multicolumn{6}{c}{DMPC} \\
  \cmidrule(lr){2-7}
  & N & P & G & C & W & average \\
\midrule
DPPC-optimised                & 3.84 & 4.61  & 8.59  & 4.94  & 6.81 & 5.76 (1.85\%) \\
DMPC-optimised                & 4.28 & 4.15  & 7.62  & 5.49  & 6.51 & 5.61 (1.81\%) \\
DOPC-optimised                & 5.89 & 8.81  & 13.44 & 8.29  & 7.74 & 8.83 (2.87\%) \\
DSPC-optimised                & 5.60 & 7.90  & 11.63 & 6.51  & 6.29 & 7.58 (2.45\%) \\
reference~\cite{Nicolia2011JCTC} & 8.53 & 10.54 & 13.32 & 10.00 & 14.64 & 11.41 (3.63\%) \\[10pt]
          & \multicolumn{6}{c}{DOPC} \\
          \cmidrule(lr){2-7}
          & N & P & G & C & W & average \\
    \midrule
DPPC-optimised                & 3.28  & 4.55  & 6.27  & 7.59  & 8.55  & 6.05 (2.03\%) \\
DMPC-optimised                & 3.77  & 3.61  & 5.44  & 8.87  & 7.22  & 6.78 (1.96\%) \\
DOPC-optimised                & 3.21  & 3.37  & 5.11  & 8.41  & 8.63  & 5.74 (1.95\%) \\
DSPC-optimised                & 3.21  & 2.98  & 5.25  & 8.80  & 10.58 & 6.16 (2.08\%) \\
reference~\cite{Nicolia2011JCTC} & 10.33 & 6.21  & 13.38 & 13.98 & 24.26 & 13.63 (4.79\%) \\[10pt]
          & \multicolumn{6}{c}{DSPC} \\
          \cmidrule(lr){2-7}
          & N & P & G & C & W & average \\
    \midrule
DPPC-optimised                & 4.24 & 3.98  & 5.13  & 6.56  & 11.04 & 6.19 (1.86\%) \\
DMPC-optimised                & 4.40 & 3.52  & 4.55  & 7.25  & 11.59 & 6.26 (1.88\%) \\
DOPC-optimised                & 4.90 & 4.03  & 5.06  & 7.36  & 10.82 & 6.43 (1.94\%) \\
DSPC-optimised                & 4.45 & 3.38  & 4.17  & 6.75  & 11.22 & 5.99 (1.80\%) \\
reference~\cite{Nicolia2011JCTC} & 8.60 & 10.30 & 11.52 & 10.85 & 22.62 & 12.78 (3.80\%) \\
\bottomrule
\end{tabular}}
\label{tab:deviations-compare-antonio}
\end{table}

The transferability of the obtained data sets is tested by performing hPF simulations for a lipid using parameters optimised on other structures. The absolute error on the density profiles obtained exchanging $\tilde\chi$ values are presented in Table~\ref{tab:deviations-compare-antonio}, and show how the global structure of the bilayers remain mostly unaffected, with relatively small changes in $S_\text{k}$ and $S_\text{p}$ values, which remain systematically lower than those of the H-F parameterisation. This fact indicates that the BO protocol is able to find robust data-sets for chemically similar moieties, also ensuring very good transferability. 
        
\clearpage
\subsection{Robustness of BO-hPF procedure}
        \begin{figure}
            \centering
            \captionsetup[subfigure]{position=bottom}
            \subfloat[]{%
                \resizebox*{7cm}{!}{\includegraphics[clip, trim=0cm 0cm 0cm 0cm]{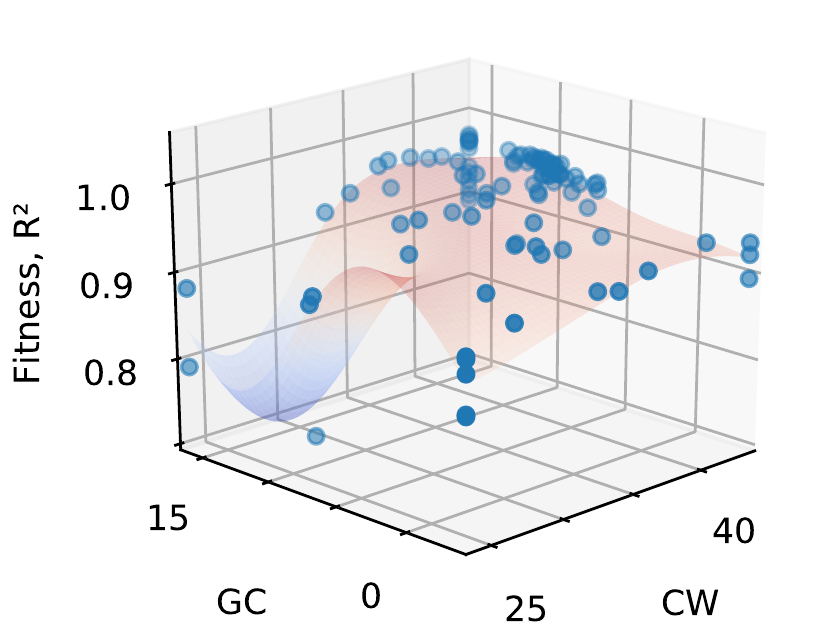}}%
            }%
            \hspace{10pt}\vrule\hspace{5pt}%
            \subfloat[]{%
                \resizebox*{6.5cm}{!}{\includegraphics{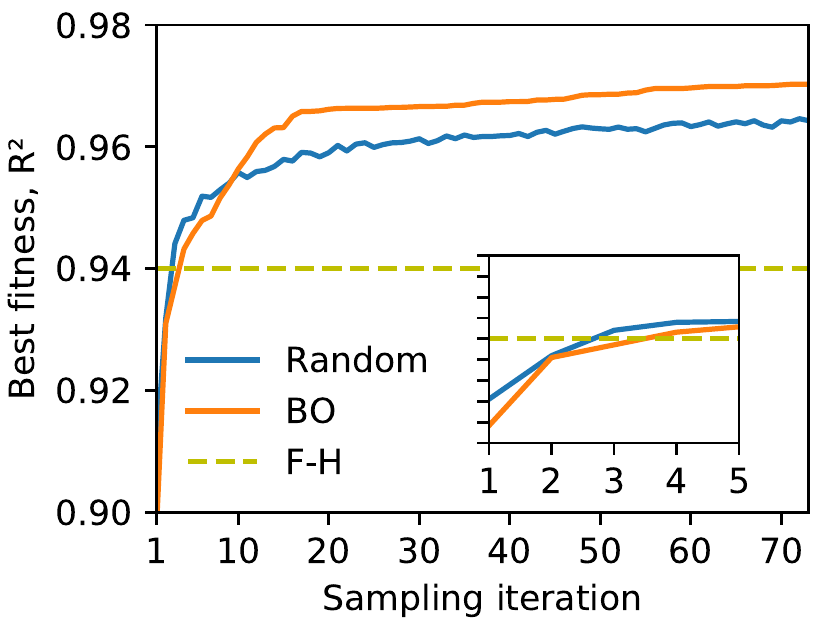}}%
            }\\
            \hrule%
            \subfloat[Bayesian optimization sampling]{%
                \resizebox*{6cm}{!}{\includegraphics{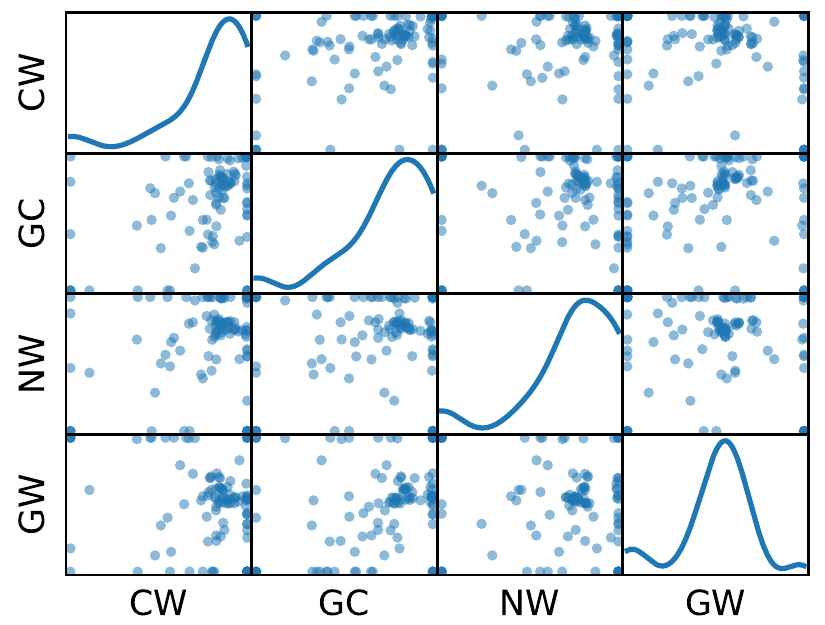}}%
            }\hspace{20pt}%
            \subfloat[Random sampling]{%
                \resizebox*{6cm}{!}{\includegraphics{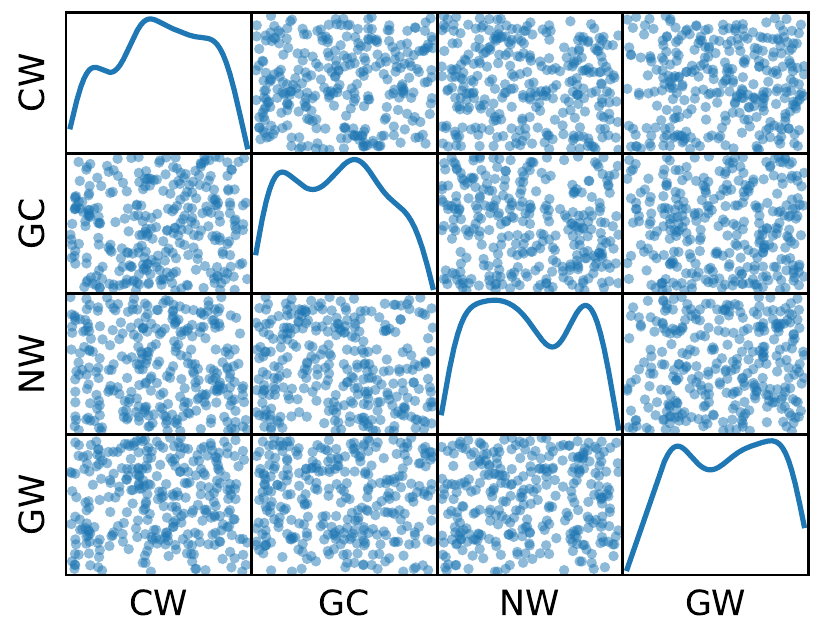}}%
            }%
            \caption{{\it Top left:} The surrogate objective fitness surface (here using the average coefficient of determination, $R^2$, across all bead species) in an example DPPC BO run with only the four parameters exhibiting the highest feature importance scores included ($\tilde{\chi}_\text{CW}$, $\tilde{\chi}_\text{GC}$, $\tilde{\chi}_\text{NW}$, and $\tilde{\chi}_\text{GW}$). Individual samplings with their associated fitnesses are represented as blue dots. A projection onto the subspace spanned by $\tilde{\chi}_\text{CW}$ and $\tilde{\chi}_\text{GC}$ shown. All $\tilde{\chi}$ matrix elements are given in \si{\kilo\joule\per\mol}.
            {\it Top right:} Best DPPC simulation membrane fitness (average $R^2$) for BO and random sampling with only the four parameters exhibiting the highest feature importance scores included. Comparison with the fitness achieved by the reference $\tilde{\chi}_\text{F-H}$ parameter set. Inset details when BO and random sampling surpass the $\tilde{\chi}_\text{F-H}$ parameter set in terms of $R^2$ fitness.
            {\it Bottom:} Scatter matrices showing correlations between all pairs of $\tilde{\chi}$ parameters in a BO run on a DPPC bilayer ({\it left}) compared with random sampling ({\it right}). Only the four parameters exhibiting the highest feature importance scores are included in the sampling. The matrix diagonal shows the density of sampled points for each individual $\tilde{\chi}_{ij}$ parameter.  All $\tilde{\chi}$ matrix elements are given in \si{\kilo\joule\per\mol}.} \label{fig:scatter-matrix}
        \end{figure}

Large multidimensional parameter spaces often exhibit multiple locally optimal parameter sets or flat fitness surfaces that can hinder convergence towards the globally optimal parameter set. Figure~\ref{fig:scatter-matrix} shows one such example for the hPF parameters, with a projection of the fitness (estimated by the surrogate fitting function) in terms of $\tilde{\chi}_\text{CW}$ and $\tilde{\chi}_\text{GC}$. The plot exhibits a narrow region of unacceptable values, and a relatively large flat plateau of high score, were the determination of the position of the maximum is numerically non trivial, and may lead to multiple solutions.

However, our tests on transferability across lipid species do not indicate such problems for the BO-hPF procedure, finding instead systematically consistent parameter sets for the different lipid species. Moreover, as shown in Figure~\ref{fig:scatter-matrix}, BO converges steadily, and outperforms random sampling protocols in finding the optimal solution, even as both schemes improve upon $\tilde{\chi}_\text{F-H}$ after only a handful of iterations. In particular, after a few efficient initial steps, random sampling is not able to converge toward the best solution, and remains confined in the large basin comprising of very different, not fully optimised, combinations of parameters. This is in agreement with results reported in the literature for BO applied to toy model functions~\cite{snoek2012practical}, and such diverse fields as e.g. chemical design~\cite{griffiths2017constrained}, active learning~\cite{brochu2010tutorial}, robotics~\cite{calandra2016bayesian,lizotte2007automatic}, and machine learning~\cite{klein2017fast}. The faster and more robust convergence of BO is determined by the intrinsic ability of the algorithm to learn what region of the space is more relevant to sample, disregarding other less relevant regions (Figure~\ref{fig:scatter-matrix}). 

    \section{Concluding remarks}
    In this work, we proposed a protocol to determine accurate potentials for hPF simulations, using BO as the main driver for the optimisation of the free parameters. Our scheme requires the definition of an arbitrary fitting function based on any set of relevant observables to be learned. The quantities of relevance may come from experimental data or from benchmark accurate higher-resolution simulations (for example all-atom or CG), the only requirement being that the pertinent quantities can be straightforwardly estimated with a hPF model.  
    
    Using DPPC, DMPC, DSPC, and DOPC phospholipid bilayers as test systems, we showed how such procedure determines sets of parameters for the interaction energy that significantly improve the models present in the literature based on F-H theory. The new Bayesian-optimised potentials also show excellent transferability among chemically similar moieties.
    
    Despite being more complex than F-H, the BO procedure here introduced offers various advantages. First, the procedure does not require the estimate of two-body interaction energies, which may be difficult to determine with good accuracy, for example, in the absence of CG models compatible with the mapping employed in the hPF simulations. Second, the protocol is very general, and can thus be used to concomitantly optimise the mixing terms of the interaction energy ($\tilde\chi$) and any other parameter of relevance present in other parts of the energy functional. This is particularly interesting in the view of recent advances for hPF model potentials, which include, for example, specific potentials for peptides, for electrostatics~\cite{Zhu2016PCCP,kolli2018JCTC,Bore2019JCTC}, or for surface energy terms~\cite{sgouros2018mesoscopic,bore2020hybrid}. Finally, being an automatic procedure, BO does not require user-based fine tuning of the parameters, ensuring more a more systematic and reproducible determination of the potentials, especially for chemically complex systems.
    
    BO is robust in determining physically meaningful parameters despite the relatively large variable space. This is due to the ability of BO to restrain the search only in a sub-region of the space where the physical solution is contained. Nonetheless, this evidence cannot be assumed as general, and it cannot be excluded that BO of hPF parameters over even higher-dimensional variable spaces would lead to numerical ambiguities. In this respect, we may suggest that the best strategy for the optimisation of hPF parameters implies the formulation of an adequate Ansatz, for example using the F-H method, that would be used as a starting point for the optimisation. In this work, we showed how feature importance can be applied to the BO procedure to identify on-the-fly those parameters that are not relevant for the convergence to the best solution, and which can be thus dropped out of the optimisation protocol. In this way,  full BO optimisation can be performed only on a subset of relevant parameters, keeping all the other at (or in the neighbourhood of) their initial F-H values. In case the F-H parameters cannot be determined, or the parameter space is intrinsically too large, we foresee the possibility of introducing  penalty terms to the fitting function, similarly to those used in other optimisation procedures like RESP~\cite{bayly1993well}, even though this has not been explored in this work.   
    
    In conclusion, the establishment of an automated machine-learning procedure for the optimisation of hPF parameters promises to further expand the applicability of such powerful simulation method toward increasingly chemically complex systems.

\section*{Acknowledgement}
    The authors thank Antonio De Nicola for providing topology and structure files for the lipid bilayer systems.
    
\section*{Data availability statement}
The data that support the findings of this study are available from the corresponding author upon reasonable request.

\section*{Disclosure statement}
    The authors declare no competing financial interest.

\section*{Funding}
    The authors acknowledge the support of the Norwegian Research Council through the CoE Hylleraas Centre for Quantum Molecular Sciences (Grant n. 262695), the Norwegian Supercomputing Program (NOTUR) (Grant No. NN4654K), and by the Deutsche Forschungsgemeinschaft (DFG, German Research Foundation), (project number 233630050 - TRR 146).
        
\bibliographystyle{tfo}
\bibliography{bo_lipids}
\end{document}